\documentclass[12pt,manuscript]{emulateapj}
\usepackage{epsfig}
\usepackage{apjfonts}
\newcommand{\df}{\dotfill}
\newcommand{\kms}{\ensuremath{\rm km\,s^{-1}}}
\newcommand{\ms}{\ensuremath{\rm m\,s^{-1}}}

\newcommand{\teff}{\ensuremath{T_{\rm eff}}}
\newcommand{\logg}{\ensuremath{\log{g}}}
\newcommand{\vsini}{\ensuremath{v \sin{i}}}

\newcommand{\mh}{\rm{[m/H]}}

\newcommand{\rsun}{\ensuremath{R_\sun}}
\newcommand{\msun}{\ensuremath{M_\sun}}

\newcommand{\rstar}{\ensuremath{R_\star}}
\newcommand{\mstar}{\ensuremath{M_\star}}
\newcommand{\loggstar}{\ensuremath{\logg_\star}}

\newcommand{\rhostar}{\ensuremath{\rho_\star}}

\newcommand{\rpl}{\ensuremath{R_{\rm P}}}
\newcommand{\mpl}{\ensuremath{M_{\rm P}}}

\newcommand{\rjup}{\ensuremath{R_{\rm J}}}
\newcommand{\mjup}{\ensuremath{M_{\rm J}}}

\shortauthors{Bryan et al.}
\shorttitle{Qatar-2}

\begin{document}

\title{Qatar-2: A K dwarf orbited by a transiting hot Jupiter and a
more massive companion in an outer orbit}


\author{
Marta~L.~Bryan\altaffilmark{1},
Khalid~A.~Alsubai\altaffilmark{2},
David~W.~Latham\altaffilmark{3},
Neil~R.~Parley\altaffilmark{4},
Andrew~Collier~Cameron\altaffilmark{4},
Samuel~N.~Quinn\altaffilmark{3},
Joshua~A.~Carter\altaffilmark{3,12},
Benjamin~J.~Fulton\altaffilmark{5},
Perry Berlind\altaffilmark{3},
Warren~R.~Brown\altaffilmark{3},
Lars~A.~Buchhave\altaffilmark{6,8},
Michael~L.~Calkins\altaffilmark{3},
Gilbert~A.~Esquerdo\altaffilmark{3},
G\'{a}bor F\H{u}r\'{e}sz\altaffilmark{3,14},
Uffe Gr{\aa}e~J{\o}rgensen\altaffilmark{6,8},
Keith~D.~Horne\altaffilmark{4},
Robert~P.~Stefanik\altaffilmark{3},
Rachel~A.~Street\altaffilmark{5},
Guillermo~Torres\altaffilmark{3},
Richard~G.~West\altaffilmark{7},
Martin~Dominik\altaffilmark{4,13},
Kennet~B.~W.~Harps{\o}e\altaffilmark{6,8},
Christine~Liebig\altaffilmark{4},
Sebastiano~Calchi~Novati\altaffilmark{9,10},
Davide~Ricci\altaffilmark{11},
Jesper~F.~Skottfelt\altaffilmark{6,8}
}

\altaffiltext{1}{Department of Astronomy, Harvard University,
Cambridge, MA 02138, USA}

\altaffiltext{2}{Qatar Foundation, PO BOX 5825 Doha, Qatar}

\altaffiltext{3}{Harvard-Smithsonian Center for Astrophysics, 60
Garden Street, Cambridge, MA 02138, USA}

\altaffiltext{4}{SUPA, School of Physics and Astronomy, University of
St Andrews, North Haugh, St Andrews, Fife KY16 9SS, UK}

\altaffiltext{5}{Las Cumbres Observatory Global Telescope Network,
6740 Cortona Drive, Suite 102, Goleta, CA 93117, USA}

\altaffiltext{6}{Niels Bohr Institute, University of Copenhagen,
Juliane Maries vej 30, 2100 Copenhagen, Denmark}

\altaffiltext{7}{Department of Physics and Astronomy, University of
Leicester, Leicester LE1 7RH, UK}

\altaffiltext{8}{Centre for Star and Planet Formation, Geological
Museum, {\O}ster Voldgade 5, 1350 Copenhagen, Denmark}

\altaffiltext{9}{Universit\`{a} degli Studi di Salerno, Dipartimento
di Fisica ``E.R. Caianiello'', Via S. Allende, 84081 Baronissi (SA),
Italy}

\altaffiltext{10}{INFN, Gruppo Collegato di Salerno, Sezione di Napoli,
Italy}

\altaffiltext{11}{Institut d'Astrophysique et de G\'{e}ophysique,
All\'{e}e du 6 Ao\^{u}t 17, Sart Tilman, B\^{a}t.\ B5c, 4000
Li\`{e}ge, Belgium}

\altaffiltext{12}{Hubble Fellow}

\altaffiltext{13}{Royal Society University Research Fellow}

\altaffiltext{14}{Konkoly Observatory of the Hungarian Academy of
Sciences, Budapest, Hungary}

\begin{abstract}

We report the discovery and initial characterization of Qatar-2b, a
hot Jupiter transiting a $V = 13.3$\,mag K dwarf in a circular orbit
with a short period, $P_{\rm b} = 1.34$ days.  The mass and radius of
Qatar-2b are $\mpl = 2.49\,\mjup$ and $\rpl = 1.14\,\rjup$,
respectively.  Radial-velocity monitoring of Qatar-2 over a span of
153 days revealed the presence of a second companion in an outer
orbit.  The Systemic Console yielded plausible orbits for the outer
companion, with periods on the order of a year and a companion mass of
at least several \mjup.  Thus Qatar-2 joins the short but growing list
of systems with a transiting hot Jupiter and an outer companion with a
much longer period.  This system architecture is in sharp contrast to
that found by {\it Kepler} for multi-transiting systems, which are
dominated by objects smaller than Neptune, usually with tightly spaced
orbits that must be nearly coplanar.


\end{abstract}


\section{INTRODUCTION}

Wide-angle ground-based photometric surveys, such as WASP and HATNet,
have been effective for the identification of close-in exoplanets that
transit their host stars.  Although most of the systems found by
these surveys are fainter than the stars targeted by Doppler surveys,
they are still bright enough to allow confirmation and
characterization of their transiting planets using follow-up
photometric and spectroscopic observations.

Most of the more than 150 confirmed transiting planets are hot Jupiters.
Of those found by ground-based surveys, only 12 are smaller than
Saturn, i.e.\ less than 9.4 Earth radii \citep[][as of September
2011]{Schneider:11}.  In contrast, most of the more than 1000
candidates identified by {\it Kepler} are smaller than Neptune, i.e.\ less
than 3.8 Earth radii \citep[e.g., see][]{Latham:11}.  However, it will
not be possible to determine spectroscopic orbits for the vast
majority of {\it Kepler's} small candidates, because the required velocity
precision is beyond the reach of present capabilities with instruments
such as HIRES on Keck 1 \citep[e.g., see][]{Batalha:11}.

Part of the problem is that most of the {\it Kepler} candidates are 14th
magnitude or fainter, a result of the fact that {\it Kepler} monitors a
region that is only 0.25\% of the sky.  To find all of the nearest and
brightest transiting planets, we now need to extend our photometric
surveys to cover the entire sky.  The prospects are good that such
surveys will eventually be pursued from space, with missions such as
TESS (an Explorer selected for Phase A by NASA) and PLATO (proposed to
ESA), but the earliest any of those spacecraft could be launched is
2016.

In the meantime, at least three teams are working on ground-based
photometric surveys designed to find the best systems for follow-up
studies.  An obvious strategy is to cover as much of the sky as
possible (idealy all), so that no prime systems are missed.  A
complementary strategy is to target smaller stars, where smaller
planets are easier to detect and characterize.  The MEarth project has
adopted both strategies by focusing on the coolest and smallest M
dwarfs, targeting a few thousand of the nearest and brightest examples
all over the sky. This approach has already yielded GJ 1214b
\citep{Charbonneau:09}, which lies in the ``Super-Earth'' transition
region between Neptune and the Earth and has attracted a lot of
interest because its favorable contrast with its host star allows
studies of its atmosphere \citep[e.g., see][]{Bean:11}.  The other two
efforts, the extensions of HATNet to HAT South and of WASP to the
Qatar Exoplanet Survey (QES), have chosen to improve the capability of
the more traditional strategy of a magnitude-limited wide-angle
survey. Both HAT South and QES have implemented cameras with larger
apertures, thus providing better photometric precision and lower rates
of contamination by faint stars close to the targets (enabled by the
more favorable pixel scale resulting from the longer focal lengths of
the larger cameras).  An important trade-off of this approach is that
it requires more detector pixels to cover the same area on the sky.

A practical implication of finding smaller planets is that the follow-up
Doppler spectroscopy needed to determine orbits and planetary masses
will demand access to the most capable facilities for very precise
radial velocities, such as HARPS on the European Southern
Observatory's 3.6-m telescope on la Silla in the South, and HARPS-N,
scheduled to come into operation in 2012 on the Telescopio Nazionale
Galileo (TNG) 3.6-m telescope operated for the Italian Institute of
Astrophysics (INAF) on La Palma in the North.

In this paper we report the discovery and initial characterization of
Qatar-2b, the second transiting planet from the Qatar Exoplanet Survey
to be confirmed \citep[for the first, see][]{Alsubai:11}.  Qatar-2b is
especially interesting, because our radial-velocity monitoring shows
that there is a second companion in the system with an orbital
period of about a year and a mass of at least several \mjup.

\section{QES DISCOVERY PHOTOMETRY}

The Qatar Exoplanet Survey (QES) is a wide-field photometric survey
for transiting planets.  The initial 5-camera CCD imaging system is
now deployed at a site in New Mexico and has been in operation for
more than two years.  It uses an array of five Canon lenses equiped
with $4\rm{K} \times 4\rm{K}$ CCDs to image an $11^{\circ} \times
11^{\circ}$ field on the sky simultaneously at two different pixel
scales \citep{Alsubai:11}.  The data are reduced at the University of
St. Andrews using pipeline software based on the image-subtraction
algorithm of \citet{Bramich:08}, and the data products are archived at
the University of Leicester, using the same architecture as the WASP
archive \citep{Pollacco:06}.

\begin{figure*}
\plotone{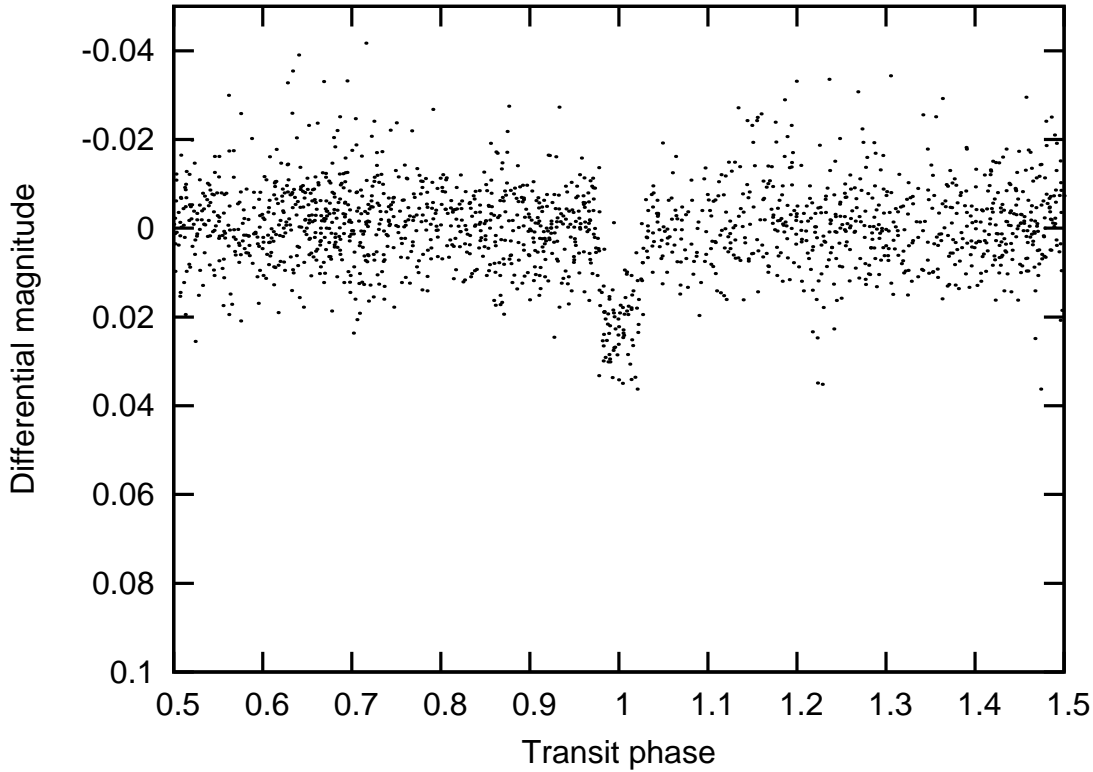}
\caption{QES discovery light curve, phased on the transit ephemeris of
Qatar-2b. The data shown here were obtained with Camera 403 in
campaigns C2, C3 and C4.}
\label{fig:lc}
\end{figure*}

The QES photometry for the $V=13.3$\,mag K dwarf that we now designate
Qatar-2 (3UC 167-129863, $\alpha_{2000}=13^{\rm h} 50^{\rm m}
37\fs41$, $\delta_{2000}=-06\arcdeg 48\arcmin 14\farcs4$
\citep{Zacharias:10}) revealed transit-like events, found by an
automated search on the archive data using the box least-squares
algorithm of \citet{Kovacs:02} as modified for the SuperWASP project
by \citet{Collier:06}.  Systematic patterns of correlated noise were
modeled and removed from the archive light curves using a combination
of the SysRem algorithm of \citet{Tamuz:05} and the trend filtering
algorithm (TFA) of \citet{Kovacs:05}.  3UC 167-129863 was screened and
selected as a promising candidate, along with two other stars in the
same QES field, using tests described by \citet{Collier:07}, designed
to ensure that the depths and durations of observed transits are
consistent with the expectations for objects of planetary dimensions
transiting main-sequence stars.  The discovery light curve, shown in
Figure \ref{fig:lc}, indicated an ephemeris of $E = 2455618.9092\pm0.0014 + N
\times 1.337098\pm0.000022$ BJD. The final photometric data for the QES light
curves are provided in full in the electronic version of the journal.

\section{FOLLOW-UP SPECTROSCOPY WITH TRES}

We monitored the spectrum of Qatar-2 using the Tillinghast Reflector
Echelle Spectrograph (TRES), mounted on the 1.5m Tillinghast Reflector
at the Fred L.~Whipple Observatory on Mount Hopkins, Arizona.  Over a
period of 153 days from 2011 January 18 to June 21 we obtained 44
spectra with a signal-to-noise ratio per resolution element (SNRe) of
at least 20 in the continuum at the center of the order containing the
Mg b features (near 518 nm).  We used the medium fiber, which yields a
resolving power of $R \sim 44,000$, corresponding to a resolution
element with $6.8\,\kms$ FWHM. The spectra were extracted using the
procedures outlined by \citet{Buchhave:10}.  The wavelength
calibration was established using exposures of a Thorium-Argon
hollow-cathode lamp illuminating the fiber, immediately before
and after the stellar observations.  A typical exposure time for a
stellar observation was 45 minutes.

\subsection{Radial Velocities}

We derived radial velocities from the TRES spectra using the
procedures for multi-order cross correlations described by
\citet{Buchhave:10}. We present here a brief summary of the technique.
First, we rejected roughly half of the echelle orders: several orders
with low SNR shortward of 446 nm; orders longward of 678 nm, which are
contaminated by telluric absorption lines or show strong fringing in
the CCD; and a few orders in between which are known to give poor
velocity performance relative to the other orders (generally because
of some other contamination such as interstellar Na D absorption or
emission lines from the Earth's atmosphere). In all, we included 23
orders in our analysis. Each observed spectrum was cross correlated,
order by order, against the corresponding orders from the strongest
single observation, the one obtained on JD 2455646. The cross
correlation functions (CCFs) from the individual orders were summed
and fit with a Gaussian function to determine the radial velocity. We
estimated the internal precision of the radial velocity for each
observation by fitting a Gaussian to the CCFs of the individual orders
and using the deviations from the global fit to calculate the standard
deviation of the mean.

We adjusted the velocities to correct for small shifts in the zero
point between observing runs, which mostly resulted from minor
modifications to the hardware between some runs.  We observed IAU
radial-velocity standard stars every night and were able to establish
the run-to-run shifts with an uncertainty on the order of $5\,\ms$.
These adjustments have been applied to the individual velocities for
Qatar-2 reported in Table \ref{tab:rvs}, which are relative to the
observation obtained on JD 2455646.  The absolute velocity of that
observed template was derived using our observations of IAU
radial-velocity standard stars and correlations of the Mg b order
against a library of synthetic templates, yielding a value of $-23.8
\pm 0.1\,\kms$ on a system where the velocity of our primary standard,
HD 182488, is defined to be $-21.508\,\kms$.  The error estimate for
this absolute velocity is fairly large, because only a single echelle
order was used in the comparison with HD~182488 by way of the
synthetic library spectra, and the absolute velocity of HD~182488
itself has an uncertainty at the level of nearly 0.1 \kms.

\begin{deluxetable}{lrrrr}
\tabletypesize{\scriptsize}
\tablewidth{0pc}
\tablecaption{Relative radial velocities of Qatar-2.\label{tab:rvs}}
\tablehead{
  \colhead{BJD} & 
  \colhead{RV\tablenotemark{a}} & 
  \colhead{$\sigma_{\rm RV}$\tablenotemark{b}} &
  \colhead{BS\tablenotemark{c}} &
  \colhead{$\sigma_{\rm BS}$} \\
  \colhead{} & 
  \colhead{(\ms)} &
  \colhead{(\ms)} &
  \colhead{(\ms)} &
  \colhead{(\ms)}\\
}
\startdata
2455580.011622 & $  231.4 $ & 33.8 & $  18.8 $ & 23.2 \\
2455581.027117 & $  383.2 $ & 37.7 & $ -31.5 $ & 22.6 \\
2455583.034587 & $ -581.6 $ & 26.7 & $  -3.8 $ & 16.8 \\
2455583.961503 & $  276.4 $ & 32.6 & $  -9.7 $ & 17.1 \\
2455585.003598 & $  327.3 $ & 27.2 & $  -6.7 $ & 13.8 \\
2455587.983438 & $  271.2 $ & 27.4 & $ -15.0 $ & 18.6 \\
2455602.958910 & $ -396.5 $ & 32.9 & $  12.2 $ & 16.0 \\
2455604.034894 & $  210.7 $ & 29.6 & $   2.7 $ & 21.0 \\
2455604.966229 & $   59.3 $ & 29.2 & $   5.3 $ & 25.4 \\
2455605.966418 & $ -737.8 $ & 23.6 & $  23.7 $ & 12.3 \\
2455607.001348 & $ -536.2 $ & 32.9 & $ -22.0 $ & 18.0 \\
2455607.987854 & $  346.7 $ & 35.5 & $  15.8 $ & 25.2 \\
2455608.955271 & $  -93.6 $ & 27.7 & $ -21.6 $ & 14.9 \\
2455610.996605 & $ -447.0 $ & 36.5 & $   4.2 $ & 25.2 \\
2455615.938163 & $  285.1 $ & 23.9 & $   3.8 $ & 15.8 \\
2455616.973605 & $  -71.0 $ & 28.9 & $ -10.2 $ & 23.1 \\
2455617.978176 & $ -799.2 $ & 21.0 & $ -30.7 $ & 13.6 \\
2455643.881697 & $  214.0 $ & 29.1 & $  19.7 $ & 17.1 \\
2455644.849974 & $ -638.7 $ & 25.5 & $  14.6 $ & 12.6 \\
2455645.901592 & $ -802.6 $ & 20.6 & $  -4.8 $ & 20.6 \\
2455646.846694 & $    0.0 $ & 17.4 & $ -28.8 $ & 11.6 \\
2455647.890953 & $  160.3 $ & 35.4 & $ -16.8 $ & 19.1 \\
2455650.857343 & $   57.0 $ & 27.1 & $   2.1 $ & 15.3 \\
2455652.895515 & $ -603.6 $ & 32.9 & $  -9.2 $ & 19.4 \\
2455656.870962 & $ -711.0 $ & 22.2 & $ -25.0 $ & 10.9 \\
2455659.877783 & $  119.4 $ & 25.4 & $  -0.8 $ & 17.5 \\
2455662.839965 & $  171.5 $ & 29.1 & $  -0.4 $ & 21.5 \\
2455663.812865 & $  -83.3 $ & 21.3 & $  -7.8 $ & 15.6 \\
2455664.867650 & $ -764.9 $ & 30.3 & $   5.6 $ & 21.7 \\
2455665.796571 & $ -549.1 $ & 21.3 & $   2.3 $ & 14.6 \\
2455668.845188 & $ -835.7 $ & 30.1 & $ 158.4 $ & 48.5 \\
2455671.762510 & $ -211.4 $ & 21.8 & $ 111.4 $ & 36.9 \\
2455672.733966 & $ -850.1 $ & 28.6 & $  26.5 $ & 13.1 \\
2455673.784772 & $ -471.2 $ & 19.2 & $ -16.2 $ & 14.9 \\
2455691.770974 & $ -292.0 $ & 32.0 & $  13.4 $ & 20.3 \\
2455702.699240 & $  229.8 $ & 24.2 & $ -22.9 $ & 19.0 \\
2455703.727260 & $ -471.1 $ & 32.2 & $ -22.6 $ & 17.6 \\
2455704.730193 & $ -724.7 $ & 19.5 & $ -20.6 $ & 12.5 \\
2455705.694921 & $  148.7 $ & 42.0 & $ -13.7 $ & 28.5 \\
2455706.738098 & $  258.4 $ & 27.6 & $ -15.8 $ & 15.8 \\
2455722.744329 & $  334.0 $ & 33.7 & $   4.6 $ & 25.0 \\
2455726.718494 & $  269.1 $ & 30.1 & $ -16.9 $ & 23.6 \\
2455728.755461 & $ -500.9 $ & 31.5 & $ -10.7 $ & 22.1 \\
2455733.706680 & $  407.0 $ & 27.5 & $ -60.7 $ & 28.9 \\

\enddata
\tablenotetext{a}{
The zero-point of these velocities is relative to the observation
obtained on BJD 2455646. The absolute velocity of that observation on
the IAU system is $-23.8 \pm 0.1\,\kms$
 }
\tablenotetext{b}{
Internal errors, summed in quadrature with the uncertainty of the
run-to-run zero point shifts, assumed to be $5\,\ms$.
}
\tablenotetext{c}{
The zero point of the bisector spans is arbitrary.
}
\end{deluxetable}

\subsection {Orbital Solution}

\begin{figure*}
\plotone{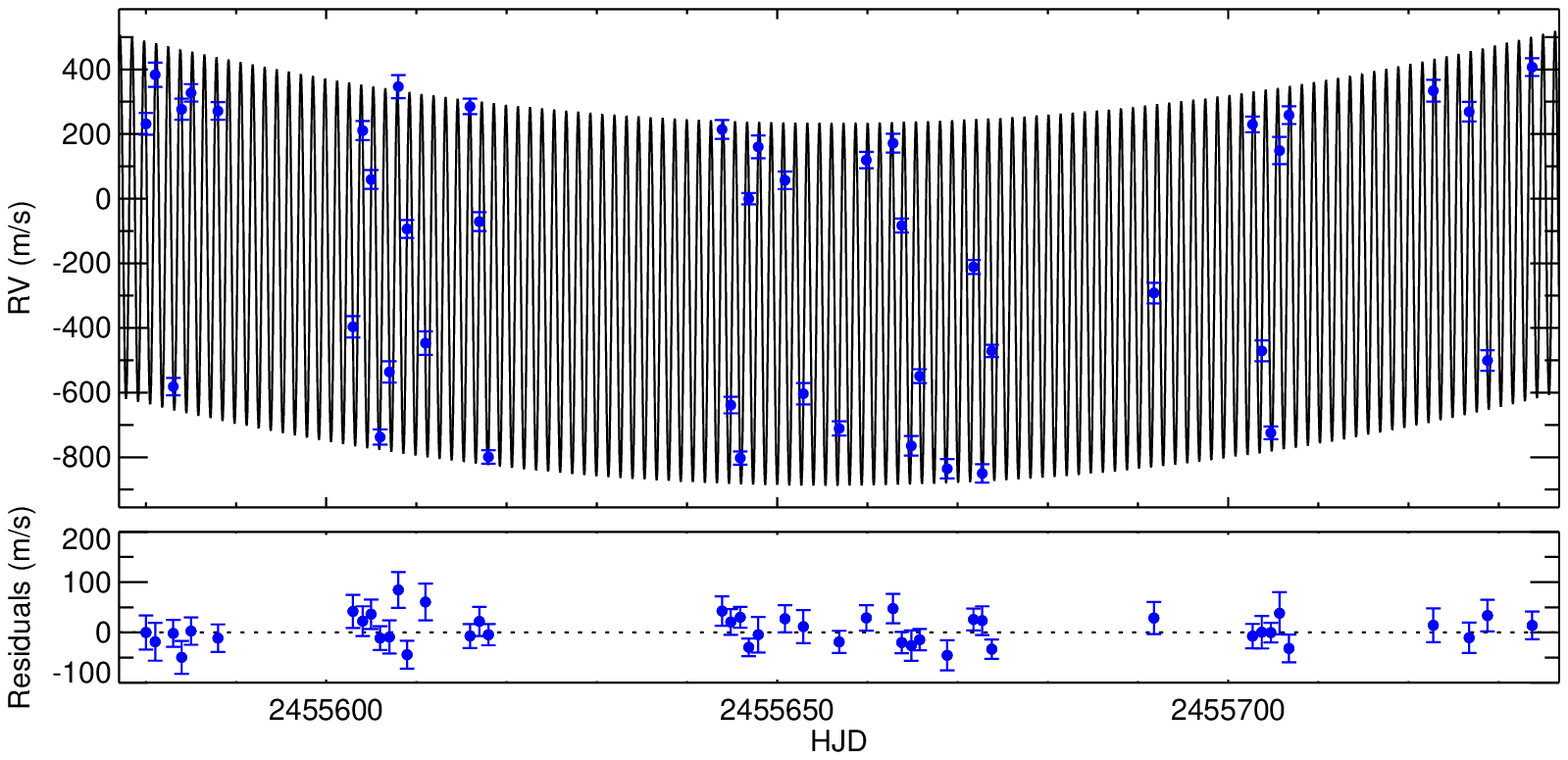}
\plottwo{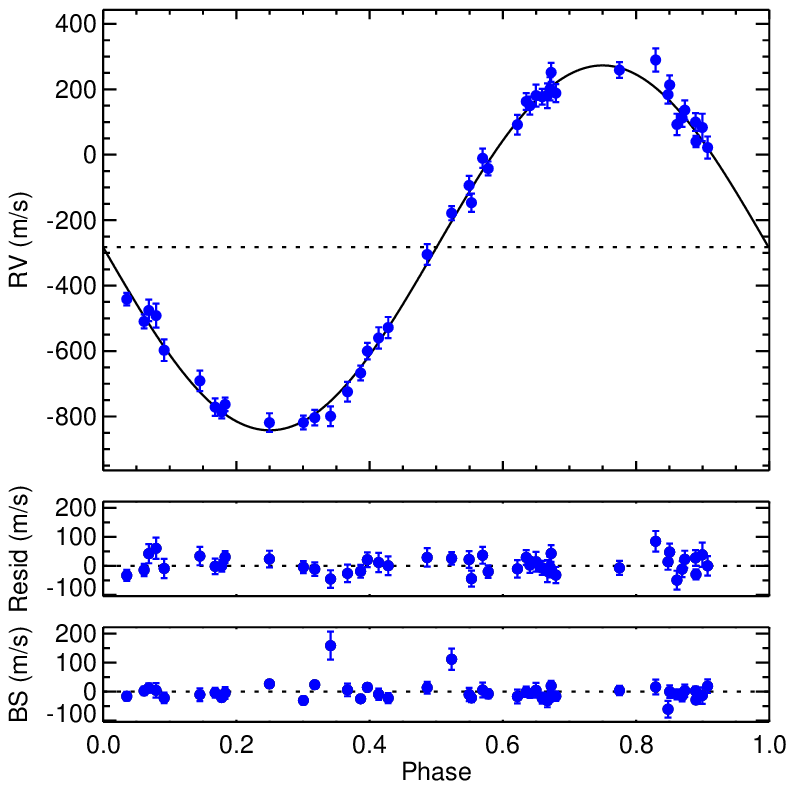}{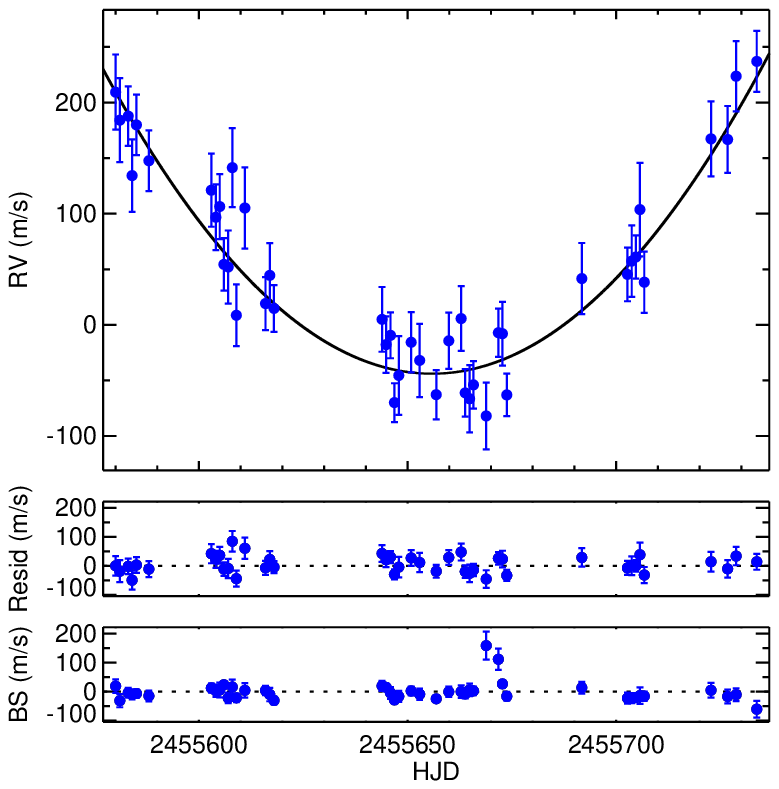}

\caption{The velocity curve for the orbit of Qatar-2b (on the bottom
  left, phased to the period from the orbital fit) and for the
  quadratic residuals due to Qatar-2c (on the bottom right), together
  with the corresponding observed velocities.  The top panel shows the
  original velocities together with the combined velocity model.  The
  panels immediately under the velocity curves show the residuals from
  the fits.  The variations in the bisector spans are plotted in the
  very bottom panels.}
\label{fig:quad}
\end{figure*}

We ran a Markov Chain Monte Carlo (MCMC) analysis on the radial
velocity data, starting with Gaussian constraints on the priors for
period and epoch from the QES discovery photometry.  We found that a
single Keplerian orbit matched the data poorly, with $\chi^2 = 424.9$
for 44 observations and 38 degrees of freedom.  Introducing two
additional terms, for a linear and quadratic velocity drift in a
Taylor expansion, resulted in a dramatic improvement to the fit, with
$\chi^2 = 46.0$ for 36 degrees of freedom (reduced $\chi^2 = 1.28$).
This model provided convincing evidence that Qatar-2 is orbited by a
transiting hot Jupiter, Qatar-2b, with period $P = 1.34$ days, plus a
second companion, Qatar-2c, with a period of roughly a year.  The
velocity curves for the orbit of Qatar-2b (phased to the period) and
the acceleration due to Qatar-2c are plotted in the bottom two
sections of Figure \ref{fig:quad}, together with the corresponding
components of the observed velocities.  The top panel shows the
original velocities together with the combined velocity model.  The
panels immediately under the velocity curves show the residuals from
the fits.

Next we used the Systemic Console \citep{Meschiari:09} to explore
possible simultaneous fits for two Keplerian orbits.  A
Levenberg-Marquardt solution gave a plausible fit and provided
starting parameters for an MCMC analyis.  When the eccentricities were
allowed to be free parameters, several plausible solutions with
similar parameters were found, all with periods a little shorter than
a year and companion masses of several \mjup.  The key parameters for
a typical solution from the Systemic Console are reported in Table
\ref{tab:orbits}, along with the parameters from the quadratic
solution.  Note that the period of the inner orbit from the radial
velocities alone, $P_{\rm RV} = 1.337169 \pm 0.000076$ is a close
match to the period from the final global solution together with all
the photometry, $P = 1.3371182 \pm 0.0000037$\,days.  A reliable
solution of the orbit of Qatar-2c will require additional
observations, which are planned for the coming season.

\begin{deluxetable}{lcc}
\tablewidth{0pc}
\tablecaption{Key Orbital Parameters \label{tab:orbits}}
\tablehead{
  \colhead{Parameter}                 &
  \colhead{Quadratic}                 &
  \colhead{Systemic\tablenotemark{a}}
}
\startdata
$P_{\rm b}$ (days)      & $ 1.337169 \pm 0.000076$  & $1.337148\pm0.000095$   \\
$K_{\rm b}$ (\ms)       & $ 559  \pm 6    $         & $559 \pm 6$             \\
$M_{\rm b}$ (\mjup)     & $ 2.48 \pm 0.03 $         & $2.49\pm0.05$           \\
$e_{\rm b}$             & $ 0.01 \pm 0.01 $         & $0.003\pm0.008$         \\
$P_{\rm c}$ (days   )   & \df                       & $332$::                 \\
$K_{\rm c}$ (\ms)       & \df                       & $301$::                 \\
$M_{\rm c}$ (\mjup)     & \df                       & $8.4$::                 \\
$e_{\rm c}$             & \df                       & $0.09$::                \\
$\Delta\gamma$\tablenotemark{b} (\ms) & $ -282.6\pm5.2  $ & \df               \\
$d\gamma/dt$  (\ms d$^{-1}$)    & $ -2.76 \pm 0.16    $ & \df                 \\
$d^2\gamma/dt^2$ (\ms d$^{-2}$) & $ 0.0875 \pm 0.0046 $ & \df                 \\
\enddata

\tablenotetext{a}{This is a typical solution from the Systemic Console;
we do not quote errors for the parameters of the outer orbit, because
the Systemic Console yielded several plausible solutions, with
differences much larger than the internal error estimates.  The
parameters given here for the outer companion should be treated with extreme
caution.}

\tablenotetext{b}{$\Delta\gamma$ is the offset of the center-of-mass
velocity for the set of relative velocities used for the orbital
solution and reported in Table \ref{tab:rvs}}

\end{deluxetable}

\subsection{Bisector Analysis}

The radial velocity of a star is defined to be the velocity of the
center of mass of the star along the line of sight.  Observationally,
radial velocities are determined by measuring the Doppler shifts of
spectral lines formed in the star's atmosphere.  Distortions of line
profiles due to phenomena such as dark spots on the surface of a
rotating star or contamination of the spectrum by another star with
variable velocity can mimic radial-velocity variations in the target
star.  A common technique for detecting line profile variations
involves measurements of line bisector spans \citep[e.g., see][]
{Queloz:01,Torres:07}.  Any variation in the bisector spans that is
correlated with variations in the radial velocities is a strong
warning that the velocity variation is probably not due to the reflex
motion induced by an orbiting planet.

The variations in the bisector spans that we measured for the 44 TRES
observations of Qatar-2 are reported in Table \ref{tab:rvs} and are
plotted in the bottom panels of Figure \ref{fig:quad}.  We were
particularly interested to see if the bisector spans showed a
correlation with the quadratic drift attributed to Qatar-2c, as might
be expected if the outer companion contributed significant light.
That would be compelling evidence that Qatar-2c was actually a star.
No such trend is apparent in the bisector spans.  There is also no
correlation with the period of Qatar-2b, supporting the planet
interpretation.

\subsection{Stellar Parameter Classification (SPC)}

The TRES spectra were also used to classify the stellar parameters of
Qatar-2 using SPC, a new tool (Buchhave et al.\ in preparation) for
comparing an observed spectrum with a library of synthetic spectra.
SPC has its origins in the procedures developed for the analysis of
spectra obtained with the CfA Digital Speedometers \citep[cf.][]
{Carney:87,Latham:02}.  It is designed to solve simultaneously for
effective temperature (\teff), metallicity (\mh), surface gravity
(\logg), projected rotational velocity (\vsini), and radial velocity
(RV), taking advantage of the higher quality and more extensive
wavelength coverage of the spectra produced by modern CCD echelle
spectrographs.  In essence, SPC
cross-correlates an observed spectrum with a library of synthetic
spectra for a grid of Kurucz model atmospheres and finds the stellar
parameters by determining the extreme of a multi-dimensional surface
fit to the peak correlation values from the grid.

For the SPC analysis of Qatar-2 we shifted the 44 observed spectra to
a common velocity and co-added them, to obtain a combined observed
spectrum with SNRe of 175 in the continuum near the Mg b features at 518 nm.
An SPC analysis yielded $\teff = 4610 \pm 50$ K, $\logg =4.65 \pm
0.10$ (cgs), $\mh = -0.02 \pm 0.08$, and $\vsini = 2.8 \pm 0.5$ \kms.
The errors quoted here are our best guesses at the limit set by the
systematic errors, which dominate the internal errors estimated from
the consistency of the results for the individual observations.  These
values of \teff\ and \mh\ are inputs to the determination of the mass
and radius of the host star (not to mention its age) using stellar
models, as described in section 6.  This is an important
issue, because the accuracy of the mass and radius determined for a
transiting planet is often limited by uncertainties in the mass and
radius of its host star.

\section{FOLLOW-UP PHOTOMETRY}

\subsection{KeplerCam Observations}

Light curves for four transits of Qatar-2b were obtained by use of
KeplerCam on the 1.2\,m telescope at the Fred Lawrence Whipple
Observatory on Mount Hopkins, Arizona.  KeplerCam utilizes a single
Fairchild 486 $4{\rm K} \times 4{\rm K}$ CCD to cover an area of
$23\arcmin \times 23\arcmin$ on the sky, with a typical FWHM of
$2\farcs5$ for stellar images.  We observed transit events of
Qatar-2b on the nights of 2011 February 26, 2011 March 6, 2011 March
14, and 2011 March 18.  The number of images captured for each of
these events was 79, 109, 94, and 168 respectively.  An SDSS $i$-band
filter was used for three of the transits, and an SDSS $g$-band filter
was used for the fourth.  For one of the $i$-band light curves we only
covered half a transit, due to increasing cloud cover around the time
of egress.  The $g$-band light curve was acquired in order to look for
possible color effects due to light from additional stars contaminating
the Qatar-2 image, such as a background eclipsing binary.

The goal of the KeplerCam observations was to produce high-quality
light curves, model them, and derive values for the radius ratio of
the planet to the star, $\rpl/\rstar$; the scaled semi-major axis of
the orbit, $a/\rstar$; and orbital inclination, $i$ (or equivalently
the impact parameter, $b$).  Differential aperture photometry was
carried out on the images, after the usual steps of bias subtraction
and flat-fielding.  We experimented with the sizes used for the
aperture for the stellar images and the annulus used for the sky
subtraction.  A smaller aperture reduces the contribution of sky
background to the noise, but risks larger systematic errors due to
imperfect centering of the stellar image. We chose the aperture and
annulus sizes that gave the best balance between these two competing
sources of error; $8\arcsec$ to $9\arcsec$ in diameter for the stellar
aperture, and an annulus between $30\arcsec$ and $60\arcsec$ for the
sky.  Sections of the out-of-transit light curves that showed
contamination by incoming clouds or dawn were rejected. Nine reference
stars were identified, and the sum of their light curves was divided
into the light curve for Qatar-2 in order to correct for atmospheric
and instrumental effects. A linear trend was then fit to the
out-of-transit sections of the resulting differential light curve and
was used to normalize the light curve for Qatar-2b to unity.

\begin{figure*}
\epsscale{0.8}
\plotone{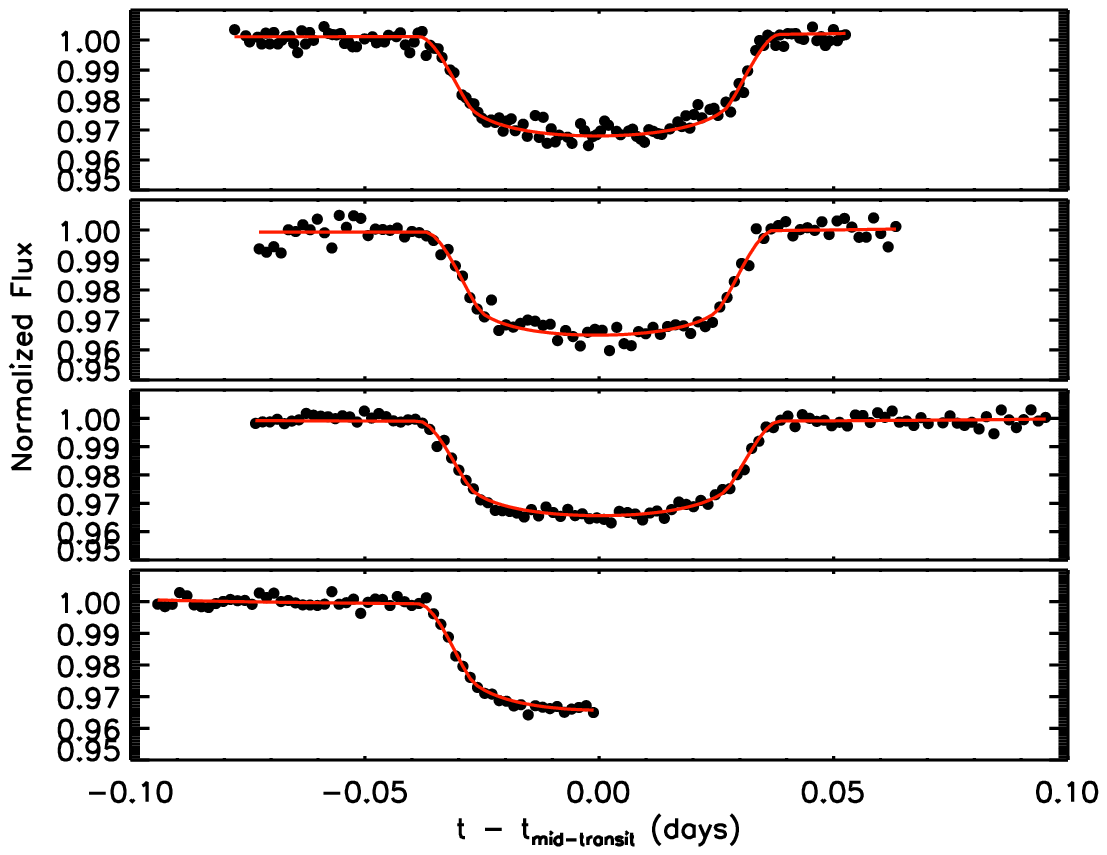}
\plotone{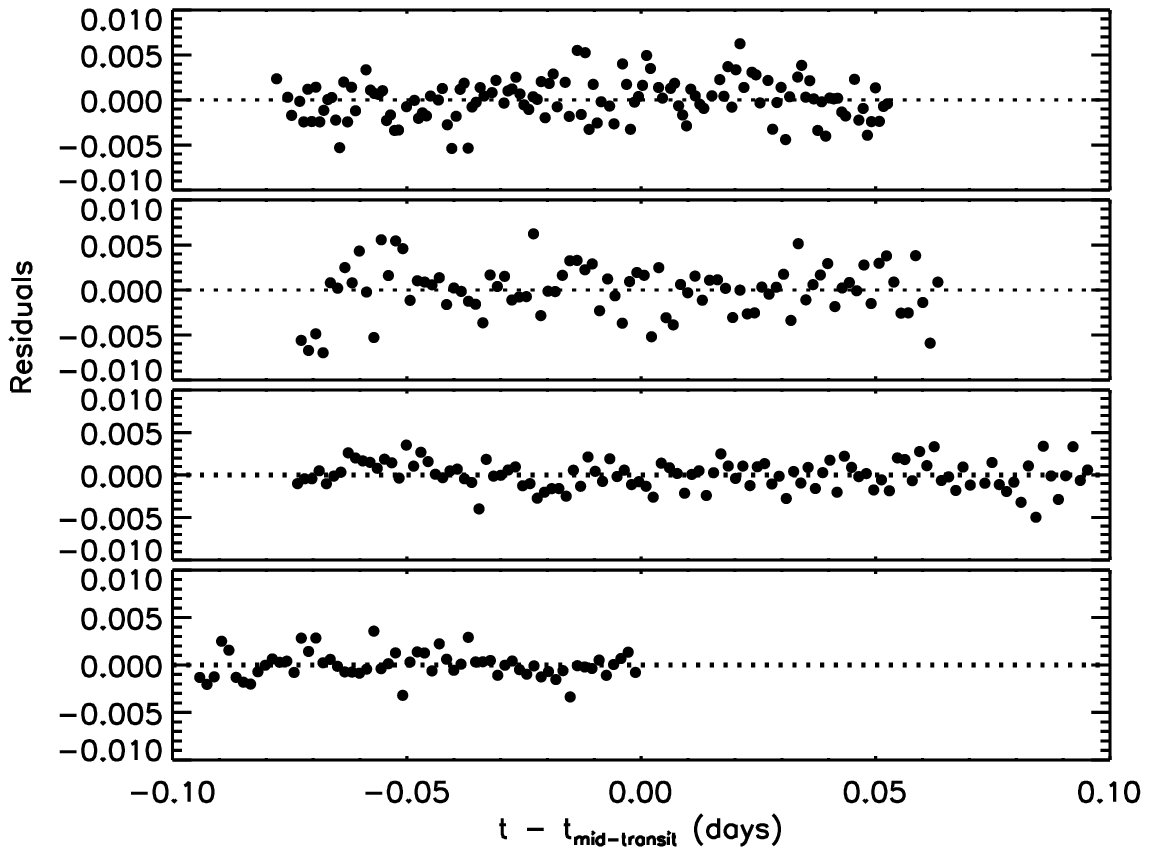}
\caption{
KeplerCam light curves for the transits of Qatar-2b. The second
light curve from the top was obtained with an SDSS $g$ filter, the other three were
obtained with an SDSS $i$ filter.  The residuals from the model are
shown underneath the light curves.  The exposure times were 120 seconds 
for the $g$ band and 60 for the $i$ band.}
\label{fig:KClightcurves}
\end{figure*}

Figure \ref{fig:KClightcurves} shows the KeplerCam light curves
together with the model fits described in section 6.
The residuals of the photometric data from the model light curves are
plotted below the light curves and show very little correlated noise.
In addition, plots of the data against column and row position on the
CCD showed no obvious correlation, and thus are not reproduced here.
The final photometric data for the KeplerCam light curves are provided
in full in the electronic version of the journal.

\subsection{LCOGT Observations}

Four partial transit events were observed with a 0.8m telescope
operated by the Las Cumbres Observatory Global Telescope (LCOGT) at
the Byrne Observatory at Sedgwick Reserve (BOS) near Santa Ynez,
CA. The Sedgwick telescope is equipped with a Santa Barbara Instrument
Group STL-6303E camera utilizing a $3{\rm K} \times 2{\rm K}$ Kodak
Enhanced KAF-6306E CCD with a pixel scale of $0.572\arcsec$ per pixel
($2 \times 2$ binning), and a $14.7\arcmin \times 9.8\arcmin$ field of
view. We observed in the SDSS $r$ band with exposure times of 120s.

The images were reduced using standard routines for bias subtraction,
dark current subtraction, and flat-field correction. We extracted
instrumental fluxes for the stellar images using PyRAF and aperture
photometry. Relative light curves were produced by dividing the fluxes
of the target star by the sum of the fluxes of four comparison stars
in each image. Each transit event was normalized to the out-of-transit
flux on that particular night. Julian dates of mid-exposure were
recorded during the observations, and later converted to BJD TDB using
the online versions of the tools described by
\citet{Eastman:10}. Aperture sizes, between $4\arcsec$ and $7\arcsec$
depending on the image quality, were chosen in order to minimize the
photometric residuals in the resulting light curves.  The final
photometric data for the LCOGT light curves are provided in full in the
electronic version of the journal.

\section{HIGH-RESOLUTION IMAGES}

A common source of astrophysical false positives for planet candidates
discovered by wide-angle photometric surveys is contamination of the
target image by an eclipsing binary, either by a physical companion in
a hierarchical system or by an accidental alignment with a background
binary.  Ground-based surveys that utilize cameras with short focal
lengths are particularly vulnerable to this problem, because the
detector pixels typically span $10\arcsec$ or more on the sky.  The
400mm focal-length Canon lenses used by the QES cameras produce images with a
typical FWHM of $7\farcs5$.  This is roughly half the size of the
images produced by the 200mm Canon lenses used by several ground-based
surveys, and thus the rate of contamination by background eclipsing
binaries should be four times better.  The images of the KeplerCam
follow-up photometry have a typical FWHM of $2\farcs5$, so this should
provide nearly an order of magnitude additional improvement in the
contamination rate.

To push to even lower limits, we obtained high-resolution images with
the Danish 1.54m telescope at the European Southern Observatory on La
Silla, by use of an Andor iXon$^{\rm EM}$+ 897 camera with an EMCCD
chip, often called a Lucky-Imaging-camera because of its ability to
obtain diffraction-limited images by recording lots of short exposures
and collecting together the best images. The camera has a pixel size
corresponding to $0\farcs09$, and was read at a frame rate of
10~Hz. In order to reach diffraction-limited images, one would usually
stack the few percentage best quality images from a sequence, but here
we have just applied the shift and add technique to all the images in
a sequence of 1,000 individual exposures, which typically will reduce
the seeing by a factor of about 3 compared to traditional CCD
observations with the same total exposure time.

\begin{figure*}
\plottwo{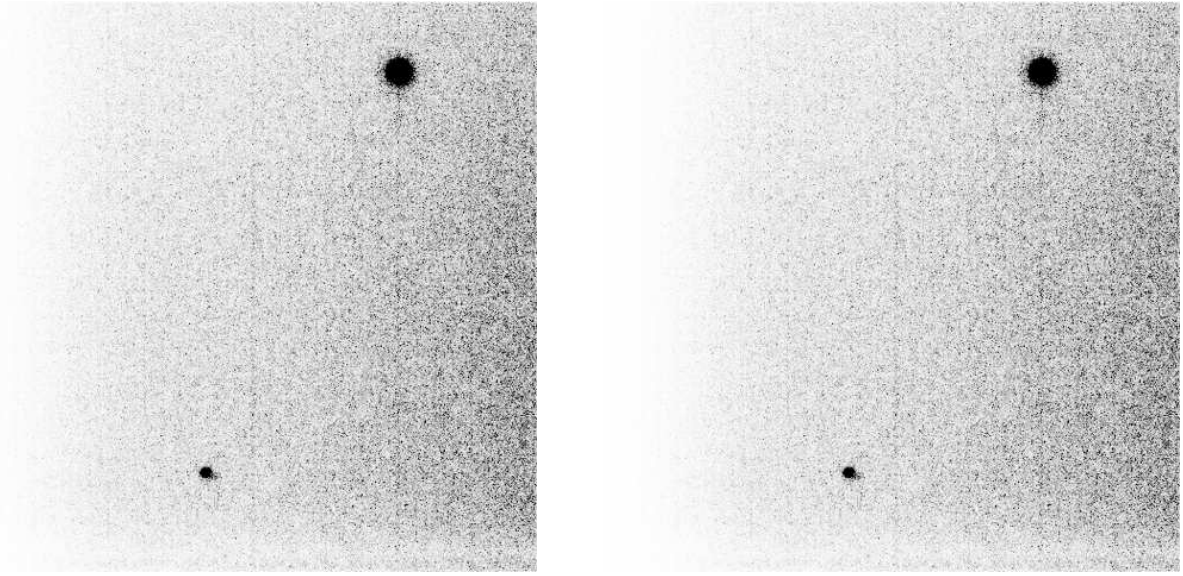}{lucky1.eps}
\caption{Lucky camera images of Qatar-2.  North is up and west is
left.  The faint pair of stars $36\arcsec$ to the south-west are
separated by slightly less than $0\farcs6$}
\label{fig:lucky}
\end{figure*}

Figure \ref{fig:lucky} shows two of the resulting images. Qatar-2 is
the bright star in the upper right part (i.e.\ north-west) of the
images. The two fainter stars in the lower part of the image
approximately $36\arcsec$ south-west of Qatar-2 are separated by
slightly less than $0.6\arcsec$.  The FWHM of the image of Qatar-2 is
about $0\farcs5$, so these images reduce the area on the sky that could be
contaminated by a background eclipsing binary by nearly another order
of magnitude compared to the KeplerCam images.  Although these images
reduce the chance of a false positive due to a background eclipsing
binary to a negligible level, they do not reduce significantly the
chance of a false positive due to contamination in a hierarchical
system, because the angular separation of most such systems is below
the resolution of these images.

\section{PHYSICAL CHARACTERISTICS OF THE STAR AND INNER PLANET
\label{sec:analysis}}

The stellar mass and radius for Qatar-2 were initially estimated using
the values of $\teff$ and $\mh$ derived from the spectra using SPC and
the value for $a/\rstar$ derived from the KeplerCam light curves,
together with isochrones from the Yonsei-Yale series of stellar models
\citep{Yi:01}, following the procedures described by
\citet{Torres:08}.  This yielded a mass of $\mstar = 0.742 \pm
0.35\,\msun$, which was then imposed as an MCMC parameter controlled
by a Gaussian prior on the simultaneous fit to the radial velocities
and all the light curves from the QES, KeplerCam, and LCOGT
photometry.  All of our exploratory fits to the radial velocities
indicated that the orbital eccentricity for Qatar-2b was
indistinguishable from circular, so in our final model we fixed that
eccentricity to zero and fit the residuals of the radial velocities
from a circular orbit using a quadratic Taylor expansion.  The transit
light curves were modeled using the formulation of \citet{Mandel:02}
in the small-planet approximation.  A four-coefficient nonlinear
limb-darkening model was used, employing fixed coefficients
appropriate to the $R$ band for the QES photometry, the SDSS $r$ band
for the LCOGT photometry, and the SDSS $i$ or $g$ band for the
KeplerCam photometry.  These coefficients were determined from the
tables of \citet{Claret:04}, interpolated to the values of \teff\ and
\mh\ determined from the TRES spectra using SPC.

\begin{figure*}
\epsscale{0.8}
\plotone{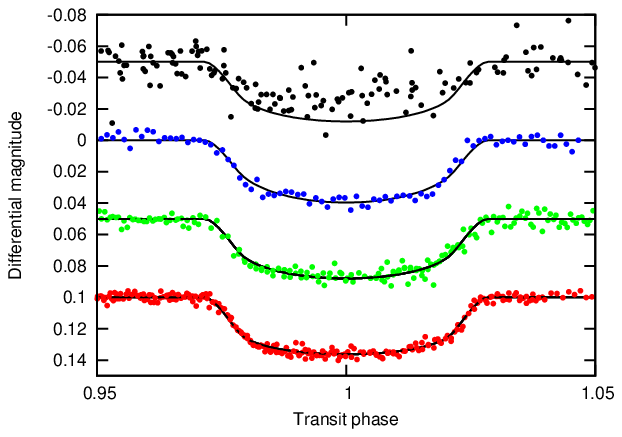}
\plotone{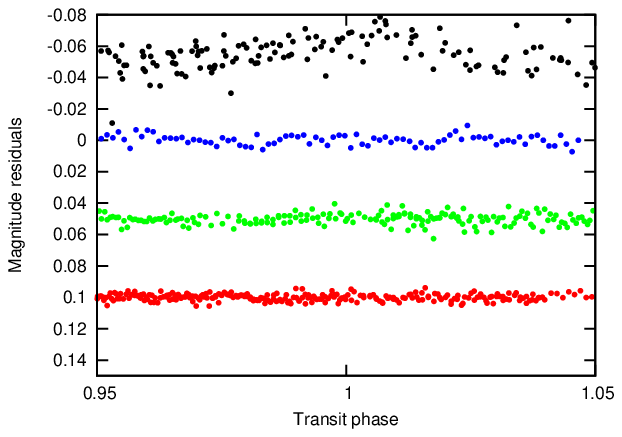}
\caption{The fit of the light curves found by the global analysis.
The first curve in the upper panel shows the QES discovery light curve
in the $R$ band; the second shows the KeplerCam SDSS $g$-band light
curve; the third shows the LCOGT SDSS $r$-band light curve, and the
fourth shows the KeplerCam SDSS $i$-band light curve.  The lower panel shows
the corresponding residuals from the global fit.  The dip in the QES 
light curve may be too shallow due to aggressive detrending.}
\label{fig:lcfits}
\end{figure*}

\begin{figure*}
\plotone{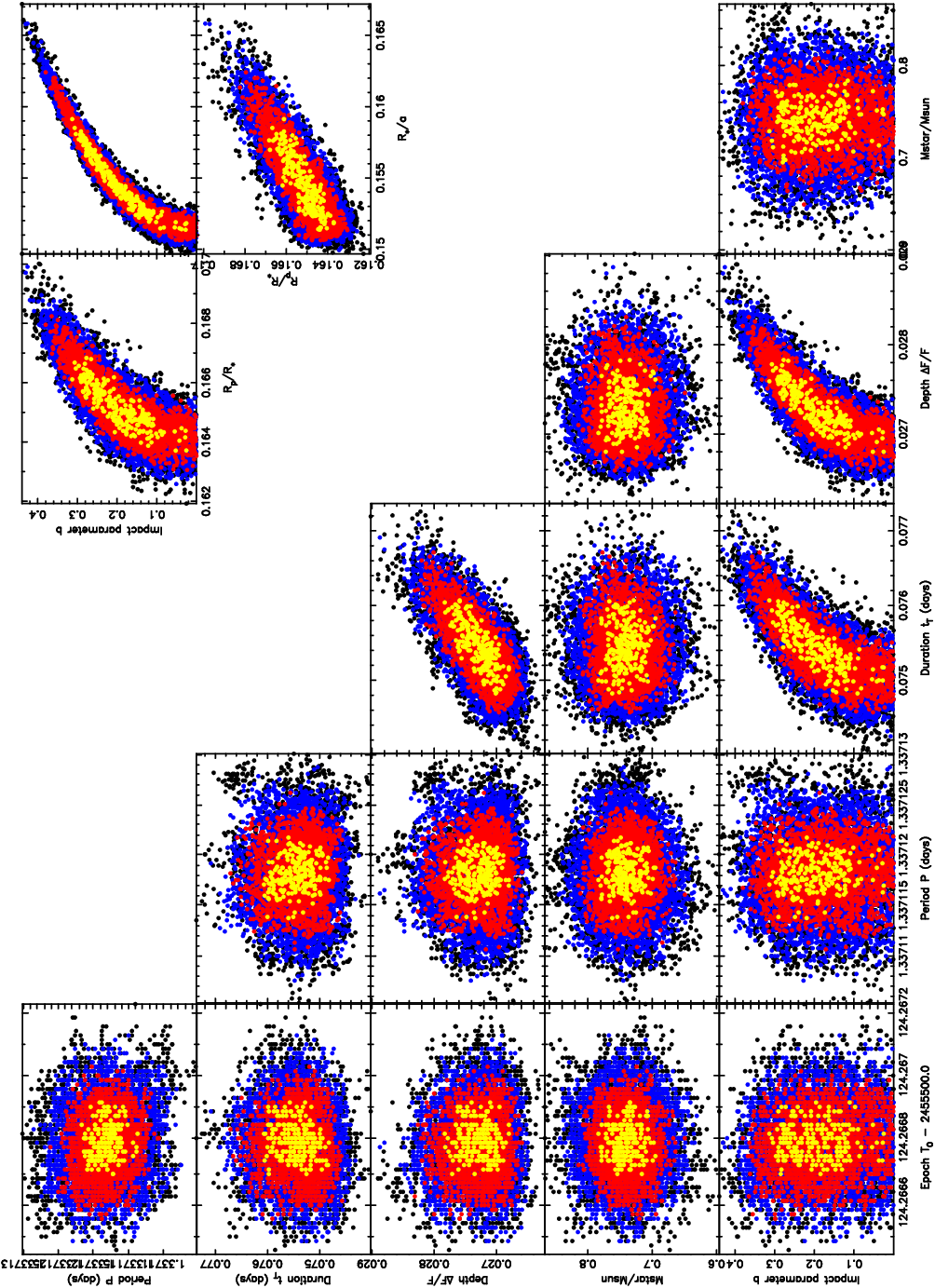}
\caption{The correlations between the various posterior
estimates from the MCMC global fit.}
\label{fig:matrix}
\end{figure*}

The parameter optimization was performed using the current version of
the MCMC code described by \citet{Collier:07} and \citet{Pollacco:08}.
The transit light curve is modeled in terms of the epoch $T_0$ of mid
transit, the orbital period $P$, the ratio of the radii squared $d =
(\rpl/\rstar)^2$, the approximate duration $t_T$ of the transit from
initial to final contact, and the impact parameter $b = (a \cos
i)/\rstar$.  The radial-velocity model is defined by the stellar
orbital velocity semi-amplitude due to the inner planet $K_b$ and
three coefficients in a quadratic Taylor expansion for the
acceleration of the star due to the outer planet, the first
coefficient being the offset of the center-of-mass velocity for the
relative velocities used for the orbital solution, $\Delta\gamma$.
The values of \teff\ and \mh\ were treated as additional MCMC
parameters, constrained by Gaussian priors with mean values and
variances as determined from the TRES spectra using SPC.  The final
light curves and corresponding fits from the global analysis are shown
in Figure \ref{fig:lcfits} for the QES discovery photometry and for
the KeplerCam and LCOGT follow-up photometry.
%
%
The correlations between the various posterior parameter estimates are
shown in the matrix of plots in Figure \ref{fig:matrix}. The final
physical and orbital parameters and error estimates for the star and
planet are reported in Table \ref{tab:results}. It is reassuring to
see the good agreement between the stellar parameters estimated from
the spectra using SPC and the final values from the global analysis:
$\teff = 4650 \pm 50$\,K from SPC compared to $4645 \pm 50$, and
$\logg = 4.65 \pm 0.10$\,cgs versus $4.601 \pm 0.018$.

\begin{deluxetable*}{llcl}
\tabletypesize{\scriptsize}
\tablewidth{0in}
\tablecaption{Planetary and Stellar Parameters.\label{tab:results}}
\tablehead{
\colhead{Parameter} &
\colhead{Symbol}    &
\colhead{Value}     &
\colhead{Units}
}
\startdata
Transit epoch (BJD TDB)           & $T_0$                 & $ 2455624.26679 \pm 0.00011 $  & days                  \\
Orbital period                    & $P_{\rm b}$           & $ 1.3371182  \pm 0.0000037  $  & days                  \\
Transit duration                  & $t_T$                 & $ 0.07540    \pm 0.00049    $  & days                  \\
Planet/star area ratio            & $(\rpl/\rstar)^2$     & $ 0.02725    \pm 0.00040    $  &                       \\
Impact parameter                  & $b$                   & $ 0.19       \pm 0.10       $  & $R_*$                 \\
Scaled stellar radius             & $\rstar/a$            & $ 0.1541     \pm 0.0030     $  &                       \\
Stellar density                   & \rhostar              & $ 2.05       \pm 0.12       $  & $\rho_\odot$          \\
Stellar effective temperature     & \teff                 & $ 4645       \pm 50         $  & K                     \\
Spectroscopic Metallicity         & \mh                   &  0 (fixed)                     &                       \\
Stellar surface gravity           & \loggstar             & $ 4.601      \pm 0.018      $  &  (cgs)                \\
Projected stellar rotation speed  & \vsini                & 2.8        (fixed)             & \kms                  \\
Stellar radius                    & \rstar                & $ 0.713      \pm 0.018      $  & \rsun                 \\
Stellar mass                      & \mstar                & $ 0.740      \pm 0.037      $  & \msun                 \\
Orbital separation                & $a$                   & $ 0.02149    \pm 0.00036    $  &  AU                   \\
Orbital inclination               & $i$                   & $ 88.30      \pm 0.94       $  & $^\circ$              \\
Stellar reflex velocity           & $K_{\rm b}$           & $ 558.7      \pm 5.9        $  & \ms                  \\
Center-of-mass velocity offset    & $\Delta\gamma$        & $ -282.6     \pm 4.0        $  & \ms                   \\
Drift in center-of-mass velocity  & $d\gamma/dt$          & $ -2.74      \pm 0.15       $  & \ms d$^{-1}$          \\
Quadratic velocity drift term     & $d^2\gamma/dt^2$      & $ 0.0870     \pm 0.0043     $  & \ms d$^{-2}$          \\
Orbital eccentricity              & $e$                   &  0 (fixed)                     &                       \\
Planet radius                     & $R_{\rm p}$           & $ 1.144      \pm 0.035      $  & \rjup                 \\
Planet mass                       & $M_{\rm p}$           & $ 2.487      \pm 0.086      $  & \mjup                 \\
Planet surface gravity            & $\log g_{\rm p}$      & $ 3.638      \pm 0.022      $  &  (cgs)                \\
Planet density                    & $\rho_{\rm p}$        & $ 1.66       \pm 0.13       $  & $\rho_{\rm J}$        \\
Planetary equilibrium temperature & $T_{\rm P}$           & $ 1292       \pm   19       $  & K                     \\
\enddata
\end{deluxetable*}

\section{DISCUSSION}

\subsection{Hot Jupiters with Companions}

One of the most important results from the {\it Kepler} mission is the
discovery of a rich population of multiple transiting planets in
systems whose orbits must be coplanar within a degree or two
\citep{Latham:11,Lissauer:11}.  Especially striking is the very low
occurrence rate for hot Jupiters in the multiple systems found by
{\it Kepler}.  In the sample of 1235 planet candidates announced by
\citet{Borucki:11}, there are 117 systems that contain a planet
candidate more massive than Saturn ($\mpl > 0.3\,\mjup$) in an orbit
with period shorter than 10 days, but only five of these systems
harbor an additional transiting planet candidate.  The remaining 112
transiting hot Jupiters are all in singles.

\begin{deluxetable*}{lccccccc}
\tabletypesize{\scriptsize}
\tablecaption{Hot Jupiters with Companions\label{tab:hj}}
\tablewidth{0pt}
\tablehead{
\colhead{Planet}         &
\colhead{\mpl (\mjup)\tablenotemark{a}}   &
\colhead{\rpl (\rjup)}   &
\colhead{$P$ (d)}        &
\colhead{$a$ (AU)}       &
\colhead{$e$}            &
\colhead{$i$ ($\degr$)}  &
\colhead{\mstar (\msun)}
}
\startdata
Qatar-2b         &$2.49\pm0.09$ &$1.14\pm0.04$ &$1.337118$   &$0.0215$ &0 (fixed)         &$88.30 \pm 0.94$    &$0.74$ \\
~~~~~~~~~~c      &$8.4$::       &\df           &$332$::      &$0.82$:: &$0.09$::          &\df                 &\df    \\
HAT-P-13b        &$0.85\pm0.04$ &$1.28\pm0.08$ &$2.916243$   &$0.0426$ &$0.014\pm0.05$    &$83.3\pm0.6$        &$1.22$ \\
~~~~~~~~~~~~~~c  &$14.5\pm1.0 $ &\df           &$448.2\pm1.0$&$1.19  $ &$0.67\pm0.02 $    &\df                 &\df    \\
HAT-P-17b        &$0.53\pm0.02$ &$1.01\pm0.03$ &$10.338523$  &$0.088 $ &$0.346\pm0.007$   &$89.2\pm0.2$        &$0.86$ \\
~~~~~~~~~~~~~~c  &$1.4^{+1.1}_{-0.4}$ &\df     & $1798^{+58}_{-89}$ &$2.8$ &$0.1^{+0.2}_{-0.1}$ &\df           &\df    \\
HAT-P-31b        &$2.2^\pm0.1 $ &$1.1\pm0.4$   &$5.005424$   &$0.055 $ &$0.245\pm0.005$   &$87.1^{+1.8}_{-2.7}$&$1.22$ \\
~~~~~~~~~~~~~~c  &$>3.4       $ &\df           &$>1022$      &\df      & \df              &\df                 &\df    \\
HD 217107b       &$1.85\pm0.05$ &\df           &$7.12689$    &$0.073 $ &$0.132\pm0.005$   &\df                 &$1.02$ \\
~~~~~~~~~~~~~~~c &$2.5\pm0.3  $ &\df           &$4210\pm190$ &$5.27  $ &$0.52\pm0.03$     &\df                 &\df    \\
HIP 14810b       &$3.9\pm0.3  $ &\df           &$6.673855 $  &$0.069 $ &$0.1427\pm0.0009$ &\df                 &$0.99$ \\
~~~~~~~~~~~~~~c  &$1.3\pm0.1  $ &\df           &$147.73  $   &$0.55  $ &$0.16\pm0.01 $    &\df                 &\df    \\
~~~~~~~~~~~~~~d  &$0.57\pm0.05$ &\df           &$962\pm15$   &$1.9   $ &$0.17\pm0.04 $    &\df                 &\df    \\
HD 187123b       &$0.52\pm0.04$ &\df           &$3.0965828   $&$0.043$ &$0.01\pm0.01 $    &\df                 &$1.06$ \\
~~~~~~~~~~~~~~~c &$2.0\pm0.3  $ &\df           &$3810\pm420$ &$4.9   $ &$0.25\pm0.03 $    &\df                 &\df    \\
$\upsilon$ Andb  &$0.69\pm0.03$ &\df           &$4.617136$   &$0.059 $ &$0.013\pm0.016$   &$>30$               &$1.27$ \\
~~~~~~~~c        &$14.0^{+2.3}_{-5.3}$ &\df    &$240.94$     &$0.83  $ &$0.245\pm0.006$   &$8\pm1$             &\df    \\
~~~~~~~~d        &$10.3^{+0.7}_{-3.3}$ &\df    &$1282$       &$2.53  $ &$0.316\pm0.006    $&$24\pm1$           &\df    \\
~~~~~~~~e        &$1.06\pm0.03$        &\df    &$3848$       &$5.25  $ &$0.0054\pm0.0004$ &\df                 &\df    \\
\enddata

\tablenotetext{a}{For the transiting hot Jupiters (the first four
entries) the actual mass is reported; for the hot Jupiters discovered
by Doppler surveys (the final four entries), the minimum mass is
listed, except for $\upsilon$ And c\&d, which are actual masses enabled by
inclinations determined with the Fine Guidance Sensors on HST.}

\end{deluxetable*}

The occurrence rate is also very low for companions to hot Jupiters
found by ground-based surveys, both photometric and spectroscopic.  A
review of the Extrasolar Planet Encyclopaedia \citep{Schneider:11}
reveals that Qatar-2b joins HAT-P-13b, HAT-P-17b, and HAT-P-31b as the
fourth transiting hot Jupiter with a confirmed outer planet, while
only four hot Jupiters found by radial velocities have outer
companions (HD 217107b, HIP 14810b, HD 187123b, and $\upsilon$
Andb). Of course, additional radial-velocity monitoring may reveal
other hot Jupiters with outer companions, and there are already hints
of velocity drifts for WASP-8 and WASP-22.  All of the confirmed
outer companions have relatively long orbital periods (see Table
\ref{tab:hj}), so it is not surprising that none of them have shown
transits, so far.  Nevertheless, it is tempting to speculate that a
close-in giant planet may stir up the orbits of other inner planets in
its system, while a system of small planets is more likely to preserve
the flatness of the disk from which it formed, allowing small planets
to survive in surprisingly compact configurations.

In only one of the systems (HIP 14810) listed in Table \ref{tab:hj}
is the outer companion less massive than the hot Jupiter.  In the
rest of the systems the nearest companions are more massive, often by
large factors.

\subsection{Orbits of Qatar-2b and HAT-P-31b}

Why is the orbit of HAT-P-31b eccentric, with $e=0.2450 \pm 0.0045$
\citep{Kipping:11}, while the orbit of Qatar-2b is indistinguishable
from circular?  One possible explanation is that both orbits started
out with significant eccentricity, perhaps as the result of a
dynamical encounter that sent each planet on a path close to its host
star, and the orbit of Qatar-2b has since been circularized by tidal
forces, while the orbit of HAT-P-31b has not.

The circularization time scale goes something like
$(\mpl/\mstar)(a/\rstar)^5$.  HAT-P-31b has a period of $P=5.00$ days
and mass of $\mpl=2.17 \mjup$, orbits a late F star with mass
$\mstar=1.22 \msun$, $a/\rstar=8.9$, and is about 3 Gyr old.  In
contrast, Qatar-2b has a much shorter period of $P_{\rm b} = 1.34$
days, similar mass of $\mpl=2.49\,\mjup$, orbits a K dwarf
with mass of $\mstar=0.74\,\msun$, $a/\rstar=6.3$, and is
very likely a member of the Galactic disk.  If Qatar-2 is three times
older than HAT-P-31, the system age divided by the circularization time
scale is nearly an order of magnitude longer for Qatar-2b, long
enough so that there has been time for tidal forces to circularize the
orbit.  In addition, the host star for HAT-P-31b is close to the mass
where the outer envelope no longer supports a convection zone, and the
circularization time scale is even longer than implied by the simple
relation that we adopted for the sake of this discussion.

Another effect that needs to be explored more carefully with dynamical
simulations is the amount of eccentricity that Qatar-2c can pump into
the orbit of Qatar-2b.  Are the tidal forces working to circularize
the orbit of Qatar-2b strong enough to keep the orbit circularized
despite the perturbations from Qatar-2c?  Finally, it would be
interesting to estimate the size and patterns of the transit time
variations expected for Qatar-2b due to perturbations by Qatar-2c.

\subsection{The Value of Deep Transits}

With a depth of about 3.5\%, the transits of Qatar-2b are deeper than
the transits of any other planet listed in The Extrasolar Planets
Encyclopaedia \citep{Schneider:11}.  The closest rival is CoRoT-2b,
with reported transit depths of 3.2\% \citep{Alonso:08}. CoRoT-2 is a
late G dwarf with $\teff=5625$ K, while Qatar-2 is a late K dwarf with
$\teff=4645$ K; the smaller radius of Qatar-2 is the main reason for
the deeper transits.  A more interesting comparison may be the bright
($V=7.7$ mag) early K dwarf HD~189733 with $\teff=5050$ K
\citep[cf.][]{Bouchy:05,Bakos:06}, which has been a favorite target
for studies of the planet's atmosphere
\citep[cf.][]{Knutson:07,Swain:08,Grillmair:08,Gibson:11}.  The
contrast between the planet and host star is even more favorable for
Qatar-2, but the system is much fainter ($V=13.3$ mag).  Thus
HD~189733 is likely to continue as a top target for follow-up studies
of hot Jupiters, although it may eventually be joined by other nearby
bright transiting systems discovered by all-sky transit surveys such
as TESS or PLATO, or by targeted searches of small cool stars such as
MEarth.  Nevertheless, Qatar-2 is a good target for amateurs and
outreach projects such as MicroObservatory
(http://mo-www.harvard.edu/MicroObservatory/) with telescopes of
modest size but covering wide fields of view, because of the
availability of many more reference stars of comparable magnitude.

\acknowledgments

G.~F. acknowledges financial support from the Hungarian OTKA-NFU
Mobility grant MB08C 81013. CL acknowledges the Qatar Foundation for
support from QNRF grant NPRP-09-476-1-078. DR (boursier FRIA)
acknowledges support from the Communaut\'e fran\c caise de Belgique -
Actions de recherche concert\'ees - Acad\'emie universitaire
Wallonie-Europe.  The Byrne Observatory at Sedgwick (BOS) is operated
by the Las Cumbres Observatory Global Telescope Network and is located
at the Sedgwick Reserve, a part of the University of California
Natural Reserve System.

{\it Facilities:} \facility{TRES}




\begin{thebibliography}{}

\bibitem[Alsubai et al(2011)]{Alsubai:11}
Alsubai, K. A., et al. 2011,
\mnras, in press (arXiv:1012.3027)

\bibitem[Alonso et al.(2008)]{Alonso:08}
Alonso, R., et al. 2008,
\aap, 482, L21

\bibitem[Bakos et al.(2006)]{Bakos:06}
Bakos, G. \'{A}., et al., 2006,
\apj, 650, 1160

\bibitem[Batalha et al.(2011)]{Batalha:11}
Batalha, N. M., et al. 2011,
\apj, 729, 27

\bibitem[Bean et al.(2011)]{Bean:11}
Bean, J. L., et al. 2011,
\apj, submitted (arXiv:1109.0582)

\bibitem[Borucki et al.(2011)]{Borucki:11}
Borucki, W. J., et al. 2011
\apj, 736, 19

\bibitem[Bouchy et al.(2005)]{Bouchy:05}
Bouchy, F., et al. 2005,
\aap, 444, L15

\bibitem[Bramich(2008)]{Bramich:08}
Bramich, D. M. 2008,
\mnras, 386, L77

\bibitem[Buchhave et al.(2010)]{Buchhave:10}
Buchhave, L. A., et al. 2010,
\apj, 720, 118

\bibitem[Carney et al.(1987)]{Carney:87}
Carney, B. W., Laird, J. B., Latham, D. W., \& Kurucz, R. L. 1987,
\aj, 93, 116

\bibitem[Charbonneau et al.(2009)]{Charbonneau:09}
Charbonneau, D., et al. 2009,
\nat, 462, 891

\bibitem[Claret(2004)]{Claret:04}
Claret, A. 2004,
\aap, 428, 1001

\bibitem[Collier Cameron et al.(2006)]{Collier:06}
Collier Cameron, A., et al. 2006,
\mnras, 373, 799

\bibitem[Collier Cameron et al.(2007)]{Collier:07}
Collier Cameron, A., et al. 2007,
\mnras, 380, 1230

\bibitem[Eastman et al.(2010)]{Eastman:10}
Eastman, J., Siverd, R., \& Gaudi, B. S. 2010,
\pasp, 122, 935

\bibitem[Gibson, Pont, \& Aigrain(2011)]{Gibson:11}
Gibson, N., Pont., F., \& Aigrain, S. 2011,
\mnras, 411, 2186

\bibitem[Grillmair et al.(2008)]{Grillmair:08}
Grillmair, C., et al. 2008,
\nat, 456, 767

\bibitem[Kipping et al.(2011)]{Kipping:11}
Kipping, D. M., et al. 2011,
\aj, 142, 95

\bibitem[Knutson et al.(2007)]{Knutson:07}
Knutson, H., et al. 2007,
\nat, 447, 183

\bibitem[Kov\'acs, Bakos, \& Noyes(2005)]{Kovacs:05}
Kov\'acs, G., Bakos, G. \'A., \& Noyes, R. W. 2005,
\mnras, 356, 557

\bibitem[Kov\'acs, Zucker, \& Mazeh(2002)]{Kovacs:02}
Kov\'acs, G., Zucker, S., \& Mazeh, T. 2002,
\aap, 391, 369

\bibitem[Latham et al.(2002)]{Latham:02}
Latham, D. W., Stefanik, R. P., Torres, G., Davis, R. J., Mazeh, T.,
Carney, B. W., Laird, J. B., \& Morse, J. A. 2002,
\aj, 124, 1144

\bibitem[Latham et al.(2011)]{Latham:11}
Latham, D. W., et al. 2011,
\apj, 732, L24

\bibitem[Lissauer et al.(2011)]{Lissauer:11}
Lissauer, J. J., et al. 2011,
\apj, accepted (arXiv1102.0543)

\bibitem[Mandel \& Agol(2002)]{Mandel:02}
Mandel, K., \& Agol, E. 2002,
\apj, 580, 171

\bibitem[Meschiari et al.(2009)]{Meschiari:09}
Meschiari, S., Wolf, A. S., Rivera, E., Laughlin, G., Vogt, S., \&
Butler, P. 2009,
\pasp, 121, 1016

\bibitem[Pollacco et al.(2006)]{Pollacco:06}
Pollacco, D. L., et al. 2006,
\pasp, 118, 1407

\bibitem[Pollacco et al.(2008)]{Pollacco:08}
Pollacco, D. L., et al. 2008,
\mnras, 385, 1576

\bibitem[Queloz et al.(2001)]{Queloz:01}
Queloz, D., et al. 2001,
\aap, 379, 279

\bibitem[Schneider(2011)]{Schneider:11}
Schneider, J. 2011,
http://exoplanet.eu (as of 24 September 2011)

\bibitem[Swain, Vasisht, \& Tinetti(2008)]{Swain:08}
Swain, M, Vasisht, G., \& Tinetti, G. 2008,
\nat, 463, 637

\bibitem[Tamuz, Mazeh, \& Zucker(2005)]{Tamuz:05}
Tamuz, O., Mazeh, T., \& Zucker, S. 2005,
\mnras, 356, 1466

\bibitem[Torres et al.(2007)]{Torres:07}
Torres, G. et al. 2007,
\apj, 666, 121

\bibitem[Torres et al.(2008)]{Torres:08}
Torres, G., Winn, J. N., \& Holman, M. J. 2008,
\apj, 677, 1324

\bibitem[Yi et al.(2001)]{Yi:01}
Yi, S. K., Demarque, P., Kim, Y.-C., Lee, Y.-W., Ree, C. H., Lejeune,
T., \& Barnes, S. 2001, 
\apjs, 136, 417

\bibitem[Zacharias et al.(2010)]{Zacharias:10}
Zacharias, N., et al. 2010,
\aj, 139, 2184


\end{thebibliography}
\end{document}